\documentclass[12pt,english]{article}
\usepackage[T1]{fontenc}
\usepackage[latin9]{inputenc}
\usepackage{float}
\usepackage{amsmath}
\usepackage{graphicx}
\usepackage{amssymb}
\usepackage{esint}
\usepackage{babel}

\begin{document}

\title{Thermal Fluctuations of the Electric Field in the Presence of Carrier
Drift}

\author{Boris Shapiro\\
 Department of Physics, Technion-Israel Institute of Technology,\\
 Haifa 32000, Israel}

\date{}
\maketitle
\begin{abstract}
We consider a semiconductor in a non-equilibrium steady state, with
a dc current. On top of the stationary carrier motion there are fluctuations.
It is shown that the stationary motion of the carriers (i.e., their
drift) can have a profound effect on the electromagnetic field fluctuations
in the bulk of the sample as well as outside it, close to the surface
(evanescent waves in the near field). The effect is particularly pronounced
near the plasma frequency. This is because drift leads to a significant
modification of the dispersion relation for the bulk and surface plasmons.
\end{abstract}

\section{Introduction}

Random thermal motion of charge carriers in a body produces a fluctuating
electromagnetic field. Properties of such a fluctuating field have
been studied for a very long time and are discussed in a number of
textbooks {[}1-4{]}. Outside the body one should distinguish between
the near and far field domain. In far field, i.e., when the distance
$\ell$ from the surface is much larger than the wavelength $\lambda$
of the corresponding Fourier component of the field, one observes
the well known phenomenon of thermal radiation. In the opposite regime,
$\ell<<\lambda$, there exists (in addition to radiation) a non-radiative
electromagnetic field that is due to evanescent waves excited by the
jiggling carriers. This random evanescent field, close to the sample
surface, is rotationless and can be described by a scalar potential.
Thermal fluctuating fields manifest themselves in a variety of experimentally
observable physical phenomena (the Casimir-Lifshitz forces, near-field
heat transfer, noncontact friction) which motivates the ongoing interest
in the subject (see {[}5-8{]} for recent reviews).

In the present paper we consider an out of equilibrium, steady state
situation when a dc current is established in a conducting medium.
Drift of the carriers can have a profound effect on the fluctuating
electromagnetic field inside, as well as outside, the medium. There
exists a large body of work on current fluctuations in the presence
of drift (for some early references see {[}9-13{]}). There is clearly
a connection between that old work and the subject of this paper,
in which we emphasize some conceptually important points and, in particular,
consider the effect of drift on the fluctuating evanescent field outside
the medium, close to its surface. It is known that drift of the carriers
affects the field outside the medium and can even lead to the existence
of a new type of weakly decaying surface wave {[}14,15{]}. The possible
effect of such waves on the electromagnetic field fluctuations close
to the sample surface was pointed out long ago {[}16{]}. Here we take
a broader view of the problem and consider a more general model of
a conducting medium. This enables us to discuss the fluctuational
field at frequencies higher than the inverse scattering time of the
carriers, when plasmonic excitations come into play. Drift is expected
to have a particularly strong effect near the plasmon frequency.

The organization of the paper is as follows: In Sec. 2 we formulate
the model and derive the corresponding permittivity tensor, with respect
to the steady state. This permittivity tensor relates quantities which
fluctuate on the background of the stationary motion of drifting carriers.
In Sec. 3 we summarize Rytov's method for treating the electromagnetic
field fluctuations and introduce modifications needed for application
of the method to our problem. In Sec. 4 we study the effect of carrier
drift on the field fluctuations well inside the sample (the limit
of an infinite medium). The experimentally more relevant case of a
field outside the sample, close to its surface, is considered in Sec.
5. It is shown there that drift of the carriers leads to a pronounced
dip in the field power spectrum in the vicinity of the bulk plasma
frequency.

\section{Permittivity tensor and plasma waves in the presence of drift}

We consider a conducting medium, e.g., a semiconductor, subject to
a constant electric field ${\bf E}_{0}$. This field causes drift
of the carriers, with the charge $e$, so that there is a steady state
current density ${\bf j}_{0}=en_{0}{\bf {v}_{0}}$, where $n_{0}$
is the equilibrium density of carriers and ${\bf v}_{0}$ is their
drift velocity. On top of the stationary motion there are fluctuations.
All fluctuating quantities will be denoted by the corresponding letters
without any subscript or superscript. For instance, ${\bf E}({\bf r},t)$
and ${\bf j}({\bf r},t)$ represent fluctuating parts of the electric
field and current density at point ${\bf r}$ and time $t$. Relations
between the fluctuating parts of various quantities are obtained by
linearization near the steady state. Since we shall be dealing with
temporary and spatial dispersion, algebraic relations (rather than
integral ones) exist only for the Fourier components. In particular,
\begin{eqnarray}
\tilde{j}_{\alpha}(\omega,{\bf k})=\sigma_{\alpha\beta}(\omega,{\bf k})\tilde{E}_{\beta}(\omega,{\bf k}),\end{eqnarray}
 where tilde indicates the Fourier transformed quantities and $\sigma_{\alpha\beta}(\omega,{\bf k})$
is the conductivity tensor with respect to the steady state (summation
over $\beta$ is implied). The specific form of this tensor depends,
of course, on the model used for the carrier motion (see below).

Let us emphasize that current density ${\bf j}$ accounts only for
the motion of the mobile carriers but not for the polarization current
of the lattice. The latter is incorporated into the dielectric constant,
$\epsilon_{L}$, of the lattice, so that the electric displacement
(its fluctuating part) is \begin{eqnarray}
\tilde{D}_{\alpha}(\omega,{\bf k})=\epsilon_{L}\tilde{E}_{\alpha}(\omega,{\bf k})+i\frac{4\pi}{\omega}\sigma_{\alpha\beta}(\omega,{\bf k})\tilde{E}_{\beta}(\omega,{\bf k})\equiv\epsilon_{\alpha\beta}(\omega,{\bf k})\tilde{E}_{\beta}(\omega,{\bf k}),\end{eqnarray}
 where $\epsilon_{\alpha\beta}(\omega,{\bf k})$ defines the steady
state permittivity tensor. We assume that the frequency $\omega$
is far from any resonant frequencies of the lattice and set $\epsilon_{L}={\rm const}$,
thus neglecting any possible dispersion effects in the lattice.

We are interested in the effect of the longitudinal plasma waves on
fluctuations. Therefore we consider a rotationless electric field,
${\bf E}({\bf r},t)=-\nabla\Phi({\bf r},t)$, which in the absence
of sources satisfies \begin{eqnarray}
\frac{\partial}{\partial x_{\alpha}}\left(\hat{\epsilon}_{\alpha\beta}\frac{\partial\Phi({\bf r},t)}{\partial x_{\beta}}\right)=0,\end{eqnarray}
 where the caret emphasizes that, in the presence of dispersion, $\hat{\epsilon}_{\alpha\beta}$
is an integral operator relating the electric displacement at point
${\bf r}$ and time $t$ to the electric field at earlier times in
some vicinity of ${\bf r}$. Fourier transforming (3) yields \begin{eqnarray}
k^{2}\epsilon(\omega,{\bf k})\tilde{\Phi}(\omega,{\bf k})=0,\end{eqnarray}
 where a scalar quantity \begin{eqnarray}
\epsilon(\omega,{\bf k})=\frac{k_{\alpha}k_{\beta}}{k^{2}}\epsilon_{\alpha\beta}(\omega,{\bf k})\end{eqnarray}
 has been introduced. The equation

\begin{eqnarray}
\epsilon(\omega,{\bf k})=0\end{eqnarray}
 defines the dispersion relation for longitudinal waves in the medium
{[}17{]}.

In order to obtain an explicit expression for $\epsilon_{\alpha\beta}(\omega,{\bf k})$
we will use a hydrodynamic equation for the carrier flow ${\bf V}$:
\begin{eqnarray}
\frac{\partial{\bf V}}{\partial t}+({\bf V}\cdot{\bf \nabla}){\bf V}=\frac{e}{m}\boldsymbol{E}-\nu{\bf V}-\frac{1}{mN}\nabla p,\end{eqnarray}
 where $m$ is the effective mass of a carrier, $\nu{\bf V}$ describes
relaxation of velocity due to collisions, with frequency $\nu$, and
the last term accounts for thermal pressure of carriers. Note that
${\bf V}$,$\boldsymbol{E}$ and $N$ refer to the total velocity,
field and carrier concentration, i.e., ${\bf V}={\bf v}_{0}+{\bf v}$,
$\boldsymbol{E}={\bf E}_{0}+{\bf E}$ and $N=n_{0}+n$. The pressure
is related to the concentration as $p=NT$, where $T$ is the temperature
in units of the Boltzmann constant $k_{B}$. 
Thermal pressure does not play a major rôle in our considerations,
which are focused on the effect of drift. It will be needed in the
treatment of the infinite medium (Sec. 5) because without thermal
pressure (or some other mechanism of spatial dispersion, e. g. diffusion)
one would encounter diverging integrals. Various versions of Eq.~(7)
are often used in semiconductor, as well as plasma, physics -- see,
e.g., Ref.~15 where the magnetic field effects are also included.
Eq.~(7) should be supplemented by the continuity equation \begin{eqnarray}
\frac{\partial N}{\partial t}+{\rm div}(N{\bf V})=0.\end{eqnarray}

Linearizing (7), (8), as well as the total current density $eN{\bf V}$,
with respect to the fluctuating quantities $n,{\bf v},{\bf E}$, and
Fourier transforming to $\omega,{\bf k}$, one obtains \begin{gather}
\tilde{{\bf j}}(\omega,{\bf k})=i\frac{e^{2}n_{0}}{m}\frac{1}{\beta+i\nu}\left\{ \tilde{{\bf E}}(\omega,{\bf k})\right.+\left({\bf k}\cdot\tilde{{\bf E}}(\omega,{\bf k})\right)\nonumber \\
\times\left.\left[\frac{1}{\beta}{\bf v}_{0}+\frac{T/m}{\beta(\beta+i\nu)-Tk^{2}/m}{\bf k}\right]\right\} ,\label{eq:1}\end{gather}
 where \begin{eqnarray}
\beta=\omega-{\bf k}\cdot{\bf v}_{0}.\end{eqnarray}
 The conductivity tensor $\sigma_{\alpha\beta}(\omega,{\bf k})$ is
readily read off from (10). We write directly the permittivity tensor,
as defined in (2): \begin{eqnarray}
\epsilon_{\alpha\beta}(\omega,{\bf k})=(\epsilon'_{L}+i\epsilon_{L}^{\prime\prime})\delta_{\alpha\beta}-\frac{\omega_{p}^{2}}{(\beta+i\nu)\omega}\left[\delta_{\alpha\beta}+\frac{1}{\beta}v_{0\alpha}k_{\beta}+\frac{Tk_{\alpha}k_{\beta}/m}{\beta(\beta+i\nu)-Tk^{2}/m}\right],\end{eqnarray}
 where $\omega_{p}^{2}=4\pi e^{2}n_{0}/m$ and the lattice dielectric
constant $\epsilon_{L}$ has been separated into the real $(\epsilon'_{L})$
and imaginary $(\epsilon_{L}^{\prime\prime})$ parts. It follows from
(5) and (11) that \begin{eqnarray}
\epsilon(\omega,{\bf k})=\epsilon'_{L}\left\{ 1-\frac{\tilde{\omega}_{p}^{2}}{(\beta+i\nu)\omega}\left[\frac{\omega}{\beta}+\frac{k^{2}R_{D}^{2}\tilde{\omega}_{p}^{2}}{\beta(\beta+i\nu)-k^{2}R_{D}^{2}\tilde{\omega}_{p}^{2}}\right]+i\frac{\epsilon_{L}^{\prime\prime}}{\epsilon'_{L}}\right\} ,\end{eqnarray}
 where $\tilde{\omega}_{p}=\omega_{p}/\sqrt{\epsilon'_{L}}$ is the
plasma frequency, renormalized by the dielectric constant of the lattice,
and $R_{D}=\sqrt{T/m\tilde{\omega}_{p}^{2}}$ is the Debye screening
radius.

Neglecting the thermal pressure term one obtains \begin{eqnarray}
\epsilon(\omega,{\bf k})=\epsilon'_{L}\left[1-\frac{\tilde{\omega}_{p}^{2}}{(\omega-{\bf k}\cdot{\bf v}_{0}+i\nu)(\omega-{\bf k}\cdot{\bf v}_{0})}+i\frac{\epsilon_{L}^{\prime\prime}}{\epsilon'_{L}}\right].\end{eqnarray}
 In equilibrium, i.e, for ${\bf v}_{0}=0$ one recovers the standard
Drude model (in the presence of the lattice): \begin{eqnarray}
\epsilon_{eq}(\omega)=\epsilon'_{L}\left[1-\frac{\tilde{\omega}_{p}^{2}}{(\omega+i\nu)\omega}+i\frac{\epsilon_{L}^{\prime\prime}}{\epsilon'_{L}}\right].\end{eqnarray}

Since plasma waves will play an important rôle in our treatment of
fluctuations, we pause to discuss briefly propagation of these waves
in the presence of drift. To see the effect of drift most clearly,
let us neglect all dissipative terms $(\nu\rightarrow0,\ \ \epsilon_{L}^{\prime\prime}\rightarrow0)$
and compare the dispersion relation in equilibrium with that for ${\bf v}_{0}\neq0$.
Using in (6) the expression (12), with $\nu=\epsilon_{L}^{\prime\prime}=v_{0}=0$
(i.e., $\beta=\omega)$, we obtain the well known dispersion relation
for the equilibrium plasma excitations: \begin{eqnarray}
k^{2}R_{D}^{2}=\frac{\omega^{2}}{\tilde{\omega}_{p}^{2}}-1,\end{eqnarray}
 which tells us that no wave can propagate for $\omega<\tilde{\omega}_{p}$.
For $\omega>\tilde{\omega}_{p}$ propagation is possible, provided
that $kR_{D}$ is a small number - otherwise the phase velocity of
the wave becomes close to the thermal velocity of the carriers and
a strong collisionless (Landau) damping sets in {[}18{]}. Thus, the
frequency of a propagating wave should be somewhat larger than the
plasma frequency, $\omega\approx\tilde{\omega}_{p}(1+\frac{1}{2}k^{2}R_{D}^{2})$.
For the out of equilibrium situation, ${\bf v}_{0}\neq0$, a similar
calculation, under the conditions $kR_{D}\ll1$, ${\bf k}\cdot{\bf v}_{0}\ll\omega$
yields the dispersion relation \begin{eqnarray}
\omega=\tilde{\omega}_{p}+{\bf k}\cdot{\bf v}_{0}+\frac{1}{2}\tilde{\omega}_{p}k^{2}R_{D}^{2},\end{eqnarray}
 so that, in contrast with the equilibrium case, propagation with
frequencies below $\tilde{\omega}_{p}$ becomes possible, if ${\bf k}\cdot{\bf v}_{0}$
is negative and sufficiently large in magnitude for (16) to be satisfied.
The necessary condition for this is $v_{0}>R_{D}\sqrt{2\tilde{\omega}_{p}(\tilde{\omega}_{p}-\omega)}$.
Although the effect of drift on plasma waves is an interesting topic
in itself, we shall not pursue it any further but rather concentrate
on the effect of drift on fluctuations.

\section{BASIC EQUATIONS OF THE THEORY}

In our treatment of fluctuations we follow Rytov's method {[}2-4{]},
in which random Langevin sources are introduced into the Maxwell equations.
These sources describe the spontaneous (thermal and quantum) fluctuations
of polarization and current density, ${\bf j}^{(s)}({\bf r},t)$.
In equilibrium the correlation function of these sources is determined
by the fluctuation-dissipation theorem and it is given in terms of
the imaginary part of the dielectric constant. Since we are dealing
with a nonequilibrium situation, there is no general relation between
the correlation function of the sources and either $\epsilon_{\alpha\beta}(\omega,k)$
or its equilibrium counterpart. However, it can happen that, in spite
of the drift of the conduction carriers, the spontaneous sources $j^{(s)}$
remain essentially the same as in equilibrium. This will occur, for
instance, when the sources $j^{(s)}$ originate primarily in the lattice,
rather than in the system of conduction carriers, i.e, when the third
term in (14) dominates over the imaginary part of the second term.
Assuming that $\nu$ is much smaller than the frequency of interest
$\omega$ and comparing the two terms in (14), one arrives at the
requirement $(\epsilon^{\prime\prime}/\epsilon'_{L})\gg\tilde{\omega}_{p}^{2}\nu/\omega^{3}$.
Under this condition the correlation function of the spontaneous random
sources is determined solely by the lattice and is given by {[}2,3{]}:
\begin{gather}
\langle j_{\alpha}^{(s)}({\bf r},\omega)j_{\beta}^{(s)^{*}}({\bf r'},\omega')\rangle=\delta_{\alpha\beta}\frac{\hbar\omega^{2}}{8\pi^{2}}\epsilon_{L}^{\prime\prime}\ {\rm coth}\ \frac{\hbar\omega}{2T}\delta({\bf r}-{\bf r}')\delta(\omega-\omega')\equiv\nonumber \\
\equiv\langle j_{\alpha}^{(s)}({\bf r})j_{\beta}^{(s)^{*}}({\bf r')\rangle_{\omega}\delta(\omega-\omega')}\label{eq2}\end{gather}
 where it was assumed that the lattice, in the presence of the conduction
current, remains close to equilibrium, at some temperature $T$. This
assumption is quite common in the transport theory of metals and semiconductors.
Thus, we arrive at a simple picture when the spontaneous fluctuations
originate in the lattice while the conduction carriers and their drift
only affect the subsequent dynamics of those fluctuations, leading
to modification of the spectral density of various fluctuating physical
quantities.

In what follows we shall be interested in the rotationless part of
the fluctuating field. It is this part that determines the evanescent
field close to the sample surface, as well as the short range correlations
in the bulk of the sample. The rotationless electric field, ${\bf E}({\bf r},t)=-\nabla\Phi({\bf r},t)$,
satisfies Eq.~(3) with added random sources: \begin{eqnarray}
\frac{\partial}{\partial x_{\alpha}}\left(\hat{\epsilon}_{\alpha\beta}\frac{\partial\Phi({\bf r},t)}{\partial x_{\beta}}\right)=-4\pi\rho^{(s)}({\bf r},t)\end{eqnarray}
 On the right hand side of (18) appears the spontaneous random charge
density, $\rho^{(s)}$, which is related to ${\bf j}^{(s)}$ by the
continuity equation, so that from (17) one obtains the expression
\begin{eqnarray}
\langle{\rho}^{(s)}({\bf r})\rho^{(s)^{*}}({\bf r}')\rangle_{\omega}=\frac{\hbar}{8\pi^{2}}\epsilon_{L}^{\prime\prime}{\rm coth}\frac{\hbar\omega}{2T}\frac{\partial^{2}}{\partial r_{\alpha}\partial r'_{\alpha}}\delta({\bf r}-{\bf r}')\end{eqnarray}
 for the spectral density. In an infinite medium (18) can be Fourier
transformed to obtain \begin{eqnarray}
k^{2}\epsilon(\omega,{\bf k})\tilde{\Phi}(\omega,{\bf k})=4\pi\tilde{\rho}^{(s)}(\omega,{\bf k})\end{eqnarray}
 where $\epsilon(\omega,{\bf k})$ is given by (12) or (13), depending
on whether one keeps or not the thermal pressure term.

Eq. (20), supplemented by the Fourier transformed Eq.~(19) \begin{eqnarray}
\langle\tilde{\rho}^{(s)}({\bf k})\tilde{\rho}^{(s)^{*}}({\bf k'})\rangle_{\omega}=\frac{\hbar k^{2}}{8\pi^{2}}(2\pi)^{3}\epsilon_{L}^{\prime\prime}{\rm coth}\frac{\hbar\omega}{2T}\delta({\bf k}-{\bf k}')\end{eqnarray}
 enables one a straightforward calculation of correlation functions
of various fluctuating quantities in an infinite medium (see Sec.
4). For finite bodies analytical treatment, in the presence of drift,
becomes difficult (see Sec. 5 for a specific example).

\section{Fluctuations in an Infinite Medium}

In this Section we consider fluctuations in the bulk of the sample,
far from the boundaries. In this case one can assume an infinite medium
and use Eqs. (20), (21), from which it immediately follows that the
spectral density for the electric potential fluctuations is \begin{eqnarray}
\langle\tilde{\Phi}({\bf k})\tilde{\Phi}^{*}({\bf k}'\rangle_{\omega}=\frac{2\hbar}{k^{2}}\frac{\epsilon_{L}^{\prime\prime}}{|\epsilon(\omega,{\bf k}|^{2}}(2\pi)^{3}{\rm coth}\frac{\hbar\omega}{2T}\delta({\bf k}-{\bf k}').\end{eqnarray}
 Multiplying this expression by the factor $k_{\alpha}k_{\beta}$
and returning to real space one obtains the spectral density of the
fluctuating electric field: \begin{eqnarray}
\langle E_{\alpha}({\bf r})E_{\beta}^{*}({\bf r}')\rangle_{\omega}=2\hbar\epsilon_{L}^{\prime\prime}{\rm coth}\frac{\hbar\omega}{2T}\int\frac{d^{3}k}{(2\pi)^{3}}\frac{k_{\alpha}k_{\beta}}{k^{2}}\frac{e^{i{\bf k}({\bf r}-{\bf r}')}}{|\epsilon(\omega,{\bf k})|^{2}}.\end{eqnarray}

Let us make a short digression to discuss the equilibrium case when
${\bf v}_{0}=0$ and $\epsilon(\omega,{\bf k})=\epsilon_{eq}(\omega)$.
In this case there is no point in making the assumption that the spontaneous
random sources originate mainly in the lattice. One can rely on the
fluctuation-dissipation theorem and use the full equilibrium dielectric
constant. For the Drude model, Eq.~(14), one arrives at the expression
\begin{eqnarray}
\langle E_{\alpha}({\bf r})E_{\beta}^{*}({\bf r}')\rangle_{\omega,{\rm equilibrium}}=2\hbar\frac{\epsilon_{{\rm eq}}^{\prime\prime}(\omega)}{|\epsilon_{{\rm eq}}(\omega)|^{2}}\ {\rm coth}\frac{\hbar\omega}{2T}\int\frac{d^{3}k}{(2\pi)^{3}}\frac{k_{\alpha}k_{\beta}}{k^{2}}e^{i{\bf k}\cdot({\bf r}-{\bf r'})}\end{eqnarray}
 where, again, the double prime denotes the imaginary part of $\epsilon_{{\rm eq}}(\omega)$.
The integral in (24) exhibits an ultraviolet divergence, indicating
a singularity for ${\bf r}'\rightarrow{\bf r}$. For instance, setting
$\beta=\alpha$ and tracing over $\alpha$ gives \begin{eqnarray}
\langle{\bf E}({\bf r})E^{*}({\bf r}')\rangle_{\omega,\ {\rm equilibrium}}=2\hbar\frac{\epsilon_{{\rm eq}}^{\prime\prime}(\omega)}{|\epsilon_{{\rm eq}}(\omega)|^{2}}\ {\rm coth}\frac{\hbar\omega}{2T}\delta({\bf r}-{\bf r}'),\end{eqnarray}
 which coincides with the second term in Eq.~(88.24) of {[}2{]} or
in Eq.~(20.26) of {[}3{]}. To remove the divergence in (24) or (25)
one must go beyond the Drude model and introduce spatial dispersion.
One source of spatial dispersion is thermal pressure which resists
any steep change of carrier concentration, thus introducing an ultraviolet
cutoff in all integrals over ${\bf k}$. All this is discussed in
detail in Ref.~3 (see also Exercise 3.12.7 in Ref.~4), where another
source of spatial dispersion, due to carrier diffusion, is also mentioned.

We now return to (23). Writing $|\epsilon(\omega,{\bf k})|^{2}=|\epsilon^{\prime}(\omega,{\bf k})|^{2}+|\epsilon^{\prime\prime}(\omega,{\bf k})|^{2}$
and assuming that the imaginary part, $\epsilon^{\prime\prime}$,
is small, one can see that the important contribution to the integral
comes from values of ${\bf k}$ in the vicinity of ${\bf k}_{c}$
which satisfies $\epsilon'(\omega,{\bf k}_{c})=0$ (the pole contribution).
But the condition $\epsilon'(\omega,{\bf k}_{c})=0$ is just the dispersion
relation for the longitudinal plasma waves, in the absence of dissipation.
The relation between plasma waves and fluctuations of the rotationless
field becomes particulary clear for the spectral densities in the
reciprocal space, such as \begin{eqnarray}
\langle E_{\alpha}E_{\beta}^{*}\rangle_{\omega{\bf k}}=2\hbar\epsilon_{L}^{\prime\prime}\ {\rm coth}\frac{\hbar\omega}{T}\:\frac{k_{\alpha}k_{\beta}}{k^{2}}\frac{1}{|\epsilon(\omega,{\bf k})|^{2}},\end{eqnarray}
 which is the Fourier transform of (23). Let us see how it works out
for $\epsilon(\omega,{\bf k})$ given in (12). We take $\nu\rightarrow0$
and use the conditions $kR_{D}<<1$, ${\bf k}\cdot{\bf v}_{0}<<\omega$.
For frequencies $\omega$ close to $\tilde{\omega}_{p}$ Eq.~(12)
simplifies to \begin{eqnarray}
\epsilon(\omega,{\bf k})=\epsilon'_{L}\left[F(\omega,{\bf k})+i\eta\right],\end{eqnarray}
 with \begin{eqnarray}
F(\omega,{\bf k})=2\frac{\omega-\tilde{\omega}_{p}-{\bf k}\cdot{\bf v}_{0}}{\tilde{\omega}_{p}}-k^{2}R_{D}^{2},\ \ \ \ \ \eta=\frac{\epsilon_{L}^{\prime\prime}}{\epsilon'_{L}}.\end{eqnarray}
 One can identify in (23) the combination $\eta/(F^{2}+\eta^{2})$
which, for small $\eta$, can be replaced by $\pi\delta(F)$ yielding
\begin{eqnarray}
\langle E_{\alpha}E_{\beta}^{*}\rangle_{\omega{\bf k}}=\pi\hbar\tilde{\omega}_{p}\ {\rm coth}\frac{\hbar\omega}{2T}\:\frac{k_{\alpha}k_{\beta}}{\epsilon'_{L}k^{2}}\delta(\omega-\tilde{\omega}_{p}-{\bf k}\cdot{\bf v}_{0}-\frac{1}{2}\tilde{\omega}_{p}k^{2}R_{D}^{2}).\end{eqnarray}
 Comparison with (16) makes it clear that contribution to the spectral
density comes only from those regions in ${\bf k}$-space where plasma
waves can propagate. Transforming (29) back into real space, one can
calculate the corresponding correlation functions.

To elucidate the effect of drift on fluctuations let us consider the
correlation function of the $x$ component of the field, i.e., $\alpha=\beta=x$,
and fix $\omega$ somewhat below $\tilde{\omega}_{p}$. Furthermore,
we chose ${\bf {v}_{0}}$ in the direction of the $x$-axis and consider
correlations along $x$-direction, taking $y=z=y'=z'=0$. Fourier
transform of (29) then yields \begin{eqnarray}
\langle E_{x}(x)E_{x}^{*}(x')\rangle_{\omega} & = & \pi\hbar\tilde{\omega}_{p}\frac{1}{\epsilon'_{L}}{\rm coth}\frac{\hbar\omega}{2T}\int\frac{d^{3}k}{(2\pi)^{3}}e^{ik_{x}(x-x')}\cdot\frac{k_{x}^{2}}{k^{2}}\\
 & \times & \delta(\omega-\tilde{\omega}_{p}-k_{x}v_{0}-\frac{1}{2}\tilde{\omega}_{p}k^{2}R_{D}^{2}),\nonumber \end{eqnarray}
 where only the essential arguments $(x,x')$ in $E_{x}$ have been
retained. In equilibrium this expression is zero, consistent with
the absence of propagating plasma waves for $\omega<\tilde{\omega}_{p}$.
Writing $k^{2}=k_{x}^{2}+q^{2}$, where ${\bf q}$ is the transverse
wave vector, and performing integration over ${\bf q}$ results in
\begin{eqnarray}
\langle E_{x}(x)E_{x}^{*}(x')\rangle_{\omega}=\frac{\hbar}{4\pi R_{D}^{2}\epsilon'_{L}}\int dk_{x}e^{ik_{x}(x-x')}\frac{k_{x}^{2}}{k_{x}^{2}+R_{D}^{-2}F(\omega,k_{x})}\theta(F(\omega,k_{x})),\end{eqnarray}
 where $F(\omega,k_{x})$ is given by Eq.~(28), with $k_{y}=k_{z}=0$,
and the step function selects the appropriate interval of $k_{x}$
in the integration region. For the integrand of (31) to be different
from zero the condition \begin{eqnarray}
v_{0}>R_{D}\sqrt{2\tilde{\omega}_{p}(\tilde{\omega}_{p}-\omega)}\end{eqnarray}
 must be satisfied. This is a necessary condition for propagation
of plasma waves below $\tilde{\omega}_{p}$. The step function in
(31) selects an interval of negative $k_{x}$ (i.e., opposite to the
direction ${\bf v}_{0})$ such that \begin{eqnarray}
\gamma-\sqrt{\gamma^{2}-2\delta}<|k_{x}|R_{D}<\gamma+\sqrt{\gamma^{2}-2\delta}\end{eqnarray}
 where \begin{eqnarray}
\gamma=\frac{v_{0}}{\tilde{\omega}_{p}R_{D}},\ \delta=1-\frac{\omega}{\tilde{\omega}_{p}}.\end{eqnarray}
 The expression (31) takes the form \begin{gather}
\langle E_{x}(x)E_{x}^{*}(x')\rangle_{\omega}=\frac{\hbar}{8\pi\epsilon'_{L}R_{D}^{3}}\intop_{\gamma-\sqrt{\gamma^{2}-2\delta}}^{\gamma+\sqrt{\gamma^{2}-2\delta}}du\exp\left(iu\frac{x-x'}{R_{D}}\right) \frac{u^{2}}{\gamma u-\delta}.\label{eq3}\end{gather}
 This example demonstrates that drift can strongly modify the fluctuation
spectrum in the vicinity of $\tilde{\omega}_{p}$ and, in particular,
can lead to emergence of fluctuations at frequencies where there were
no fluctuations in equilibrium. In addition, due to the oscillating
factor in the integrand of (35), there will be oscillations in the
field correlation function. 
Let us note that it is desirable to keep $v_{0}$ smaller than $\tilde{\omega}_{p}R_{D}$.
Indeed, $\tilde{\omega}_{p}R_{D}$ is of the order of the carrier
thermal velocity $v_{T}$. If $v_{0}$ exceeds $v_{T}$, then heating
of the carriers becomes appreciable and the hydrodynamic equation
(7) would require some modification. Thus, we shall assume $\gamma$
to be smaller than unity. 
In Fig.~1 we give an example of the spectral density in Eq.~(35),
as a function of $x-x'$, for $\gamma=0.3,\delta=0.01$. The real
part of the ratio $8\pi\epsilon'_{L}R_{D}^{3}\langle E_{x}(x)E_{x}^{*}(x')\rangle_{\omega}/\hbar{\rm coth}\frac{\hbar\omega}{2T}\equiv f(\frac{x-x'}{R_{D}})$
is plotted as a function of $\frac{x-x'}{R_{D}}$.%
\begin{figure}[H]
\begin{centering}
\includegraphics[scale=0.75]{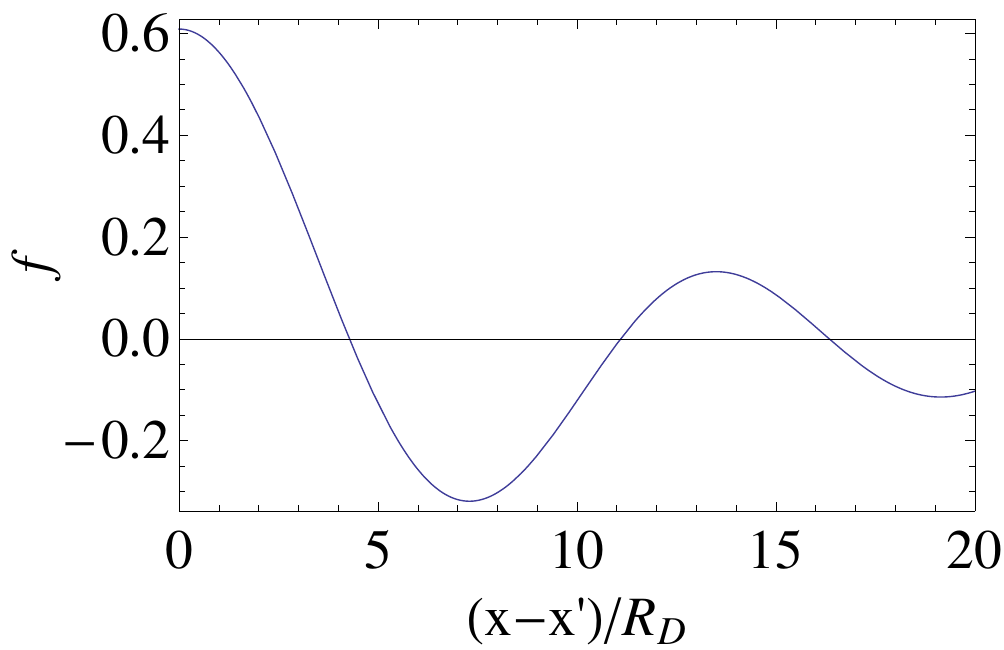}
\par\end{centering}

\caption{The real part of the normalized spectral density $\langle E_{x}(x)E_{x}^{*}(x')\rangle_{\omega}$
as a function of the distance $\frac{x-x'}{R_{D}}$.}

\end{figure}

\section{Fluctuations near the surface}

In this section we study fluctuations of the evanescent electric field
which exists close to the surface of a sample, due to fluctuating
charges and carriers inside the sample. We consider the simplest geometry
of a sample occupying half space $(z<0)$ , while the second half
$(z>0)$ is vacuum. A dc current is flowing in the medium and the
conduction carriers are drifting in the $x$-direction with velocity
$v_{0}$ (Fig.~2).%
\begin{figure}[H]
\begin{centering}
\includegraphics[scale=0.75]{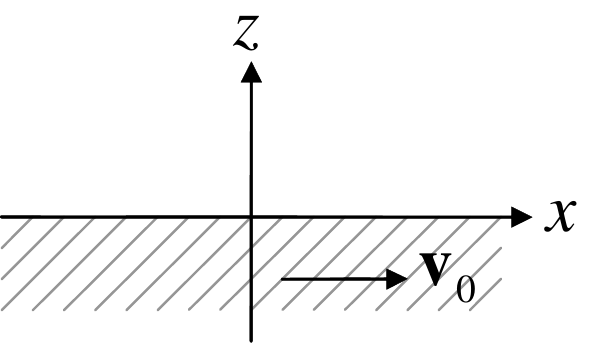}
\par\end{centering}

\caption{The sample at temperature $T$, with drifting carriers, is separated
from the vacuum $\left(T=0\right)$ by a sharp boundary (the $z=0$
plane).}

\end{figure}

Let us emphasize that in this section, unlike the previous one, the
entire system (sample + environment) is out of equilibrium already
in the absence of drift (${\bf v}_{0}=0$). It is assumed that the
sample is in local equilibrium, at temperature $T$, whereas the environment
is \textquotedbl{}cold\textquotedbl{} ($T=0$). In this case, which
is often assumed in the studies of the electromagnetic field fluctuations
{[}3,5{]}, the sample is the only source of fluctuations so that no
radiation is impinging on the sample from outside. Moreover, the zero-point
fluctuations, which exist also in the vacuum, cancel out in the process
of the electric field measurement. The latter belongs to the class
of \textquotedbl{}absorption measurements\textquotedbl{}, because
some amount of energy must be diverted into the measuring device (the
probe). This implies that instead of the symmetrized correlation function
for the random sources (Eq. (17) or (19)), with its characteristic
factor ${\rm coth}\frac{\hbar\omega}{2T}$, one should use the normally
ordered correlation function (see {[}5{]} and references therein).
The latter is obtained from its symmetrized counterpart by replacing
$\frac{1}{2}{\rm coth}\frac{\hbar\omega}{2T}$ with the Planck function
$[exp(\frac{\hbar\omega}{T})-1]^{-1}\equiv\Pi(\omega,T)$. This replacement
amounts to disregarding the zero-point fluctuations and it will be
used throughout this section {[}19{]}.

Because of the absence of translational symmetry in the $z$-direction
it is not possible anymore to Fourier transform Eq.~(18) in that
direction. This complicates the analytic treatment considerably. One
simplification, though, is that unlike the case of the infinite medium
no ultraviolet cutoff will be needed in the present geometry. Therefore
we will discard thermal pressure altogether, and use the permittivity
tensor (11) without the last term. Furthermore, we will keep the ${\bf k}$-dependence
due to drift only in the component $\epsilon_{xx}$, thus arriving
at a diagonal permittivity tensor: \begin{eqnarray}
 &  & \epsilon_{xx}=\epsilon'_{L}\left[1-\frac{\tilde{\omega}_{p}^{2}}{(\omega-k_{x}v_{0}+i\nu)(\omega-k_{x}v_{0})}+i\frac{\epsilon_{L}^{\prime\prime}}{\epsilon'_{L}}\right]\\
 &  & \epsilon_{yy}=\epsilon_{zz}=\epsilon'_{L}\left[1-\frac{\tilde{\omega}_{p}^{2}}{(\omega+i\nu)\omega}+i\frac{\epsilon_{L}^{\prime\prime}}{\epsilon'_{L}}\right]\equiv\epsilon_{0}(\omega).\nonumber \end{eqnarray}
 Eq.~(18), Fourier transformed in the $x,y$-plane, assumes the form:
\begin{eqnarray}
\epsilon_{0}(\omega)\left[-\frac{\partial^{2}}{\partial z^{2}}+\frac{\epsilon_{xx}(\omega,k_{x})}{\epsilon_{0}(\omega)}k_{x}^{2}+k_{y}^{2}\right]\tilde{\Phi}(k_{x},k_{y};z)=4\pi\tilde{\rho}^{(s)}(k_{x},k_{y};z),\end{eqnarray}
 where tilde indicates the in-plane Fourier transform. The solution
of this equation can be written in terms of a Green's function satisfying
the equation \begin{eqnarray}
\epsilon_{0}(\omega)\left(-\frac{\partial^{2}}{\partial z^{2}}+q^{2}\right)g(z,z';q)=\delta(z-z')\end{eqnarray}
 Since the right-hand-side of (37) differs from zero only inside the
medium, while the solution of interest is outside the medium, we need
$g(z,z';q)$ for $z'<0$, $z>0$. The corresponding expression can
be found, for instance, in Ref.~20: \begin{eqnarray}
g(z,z';q)=\frac{1}{(\epsilon_{0}+1)q}e^{-q(z-z')}\end{eqnarray}
 and the solution of (37) is written as \begin{eqnarray}
\tilde{\Phi}(k_{x},k_{y};z)=4\pi\int_{-\infty}^{0}dz'g(z,z';q)\tilde{\rho}^{(s)}(k_{x},k_{y};z),\end{eqnarray}
 with \begin{eqnarray}
q=\left[\frac{\epsilon_{xx}(\omega,k_{x})}{\epsilon_{0}(\omega)}k_{x}^{2}+k_{y}^{2}\right]^{1/2}.\end{eqnarray}
 The correlation function for the random sources $\tilde{\rho}^{(s)}$
follows directly from Eq.~(19), with the aforementioned replacement
of $\quad$ $\frac{1}{2}{\rm coth}\frac{\hbar\omega}{2T}$ $\quad$
by $\quad$ $\Pi(\omega,T)$ : \begin{eqnarray}
 &  & \langle\tilde{\rho}^{(s)}(k_{x},k_{y};z_{1})\tilde{\rho}^{(s)^{*}}(k'_{x},k'_{y};z_{2})\rangle_{\omega}=\\
 &  & \hbar\epsilon_{L}^{\prime\prime}\Pi(\omega,T)\delta(k_{x}-k_{x}')\delta(k_{y}-k_{y}')\left[(k_{x}^{2}+k_{y}^{2})\delta(z_{1}-z_{2})+\frac{\partial^{2}}{\partial z_{1}\partial z_{2}}\delta(z_{1}-z_{2})\right].\nonumber \end{eqnarray}
 With the help of (40), (42) and transforming back to real space,
one obtains \begin{eqnarray}
 &  & \langle\Phi(x,y,z)\Phi^{*}(x',y',z')\rangle_{\omega}=4\hbar\epsilon_{L}^{\prime\prime}\Pi(\omega,T)\int\int\frac{dk_{x}dk_{y}}{(2\pi)^{2}}e^{ik_{x}(x-x')+ik_{y}(y-y')}\times\nonumber \\
 &  & \int_{-\infty}^{0}dz_{1}\left[(k_{x}^{2}+k_{y}^{2})g(z,z_{1};q)g^{*}(z',z_{1};q)\right.\left.+\frac{\partial g(z,z_{1};q)}{\partial z_{1}}\frac{\partial g^{*}(z',z_{1};q)}{\partial z_{1}}\right].\end{eqnarray}
 From this expression one can calculate correlation functions for
various components of the electric field. We limit ourselves to the
$x$-component of the field and consider correlations in the $x$-direction,
for fixed $y,z$. The resulting function depends only on $x-x'$ and
$z$, and is given by \begin{eqnarray}
 &  & \langle E_{x}(x,y,z)E_{x}^{*}(x',y,z)\rangle_{\omega}=4\hbar\frac{\epsilon_{L}^{\prime\prime}}{|\epsilon_{0}(\omega)+1|^{2}}\Pi(\omega,T)\times\nonumber \\
 &  & ~~~~\int\int\frac{dk_{x}dk_{y}}{(2\pi)^{2}}k_{x}^{2}e^{ik_{x}(x-x')}\left(\frac{k_{x}^{2}+k_{y}^{2}}{|q|^{2}}+1\right)\frac{1}{2q'}e^{-2q'z},\end{eqnarray}
 where $q'$ is the real part of the quantity defined in Eq.~(41).

In the absence of drift, when $\epsilon_{xx}=\epsilon_{0}(\omega)$,
we have $q=\sqrt{k_{x}^{2}+k_{y}^{2}}$ and the integral in (44) can
be computed with the result \begin{gather}
\langle E_{x}(x,y,z)E_{x}^{*}(x',y,z)\rangle_{\omega}^{(0)}=\frac{\hbar}{4\pi}\frac{\epsilon_{L}^{\prime\prime}}{|\epsilon_{0}(\omega)+1|^{2}}\Pi(\omega,T)\frac{1}{z^{3}}\frac{1-\frac{X^{2}}{2z^{2}}}{(1+\frac{X^{2}}{4z^{2}})^{5/2}}\end{gather}
 where $X=|x-x'|$. Actually, more generally, one should replace in
this equation $\epsilon_{L}^{\prime\prime}$ by $\epsilon_{0}^{\prime\prime}(\omega)$.
This is because in the absence of drift there is no need to make the
additional assumption that the main source of spontaneous fluctuations
is the lattice, rather than the conduction carriers. Eq. (45) is in
agreement with the corresponding expression in {[}21{]} (our definition
of spectral density differs by a factor $2\pi$ from that in {[}21{]}).
Setting in Eq.~(45) $X=0$, one observes that the spectral power
of field fluctuations, $\langle E^{2}(z)\rangle_{\omega}$, increases
as $z^{-3}$ when the surface of the sample is approached {[}3{]}
(the divergence for $z\rightarrow0$ will be eventually cut off by
some mechanism of spatial dispersion).

It should be also noted that the factor $|\epsilon_{0}(\omega)+1|$
in (45) is due to the surface plasmon wave which appears at the frequency
satisfying the relation $Re\epsilon_{0}(\omega)=-1$. At this frequency,
and provided that dissipation is small, a strong peak appears in the
power spectrum {[}5{]}. On the other hand, the bulk plasmon frequency,
which corresponds to $Re\epsilon_{0}(\omega)=0$ (i.e., $\omega\approx\tilde{\omega}_{p})$,
does not play any special role in Eq.~(45). The frequency $\tilde{\omega}_{p}$
does become important, however, in the presence of drift, as we show
next.

Let us first take a closer look at the ratio $\epsilon_{xx}/\epsilon_{0}$
which appears in expression (41) for $q$. For $v_{0}=0$ this ratio
is unity. For $v_{0}\neq0$, however, it can become large if the frequency
$\omega$ is close to the bulk plasma frequency $\tilde{\omega}_{p}$.
Indeed, $\epsilon_{0}(\omega)$ will be then close to zero while $\epsilon_{xx}(\omega,k_{x})$
can be much larger due to finite $v_{0}$. Taking $\nu\rightarrow0$
and assuming \begin{eqnarray}
\frac{\epsilon_{L}^{\prime\prime}}{\epsilon_{L}}\ll\ |1-\frac{\tilde{\omega}_{p}}{\omega}|\ll1,\end{eqnarray}
 we have, from (36): \begin{eqnarray}
Re\frac{\epsilon_{xx}(\omega,k_{x})}{\epsilon_{0}(\omega)}=1-\frac{2(k_{x}/k_{0})-(k_{x}/k_{0})^{2}}{2\eta[1-(k_{x}/k_{0})]^{2}}\end{eqnarray}
 where \begin{eqnarray}
\eta=1-\frac{\tilde{\omega}_{p}}{\omega},\ \ k_{0}=\frac{\omega}{v_{0}}.\end{eqnarray}
 In order to estimate the integral in Eq.~(44) one has to find the
relevant values of $k_{x}$, making the main contribution to the integral.
The factor $\exp(-2q'z)$ in the integrand limits the value of $q'$
to $q'\lesssim1/z$, which in turn results in an efficient cutoff
for $k_{x}$, if $z$ is not too small. In order to make a quantitative
estimate we assume that the relevant region of $k_{x}$ corresponds
to $|k_{x}|\ll k_{0}$ and check the consistency of this assumption
later. For $|k_{x}|\ll k_{0}$ Eq.~(47) simplifies to \begin{eqnarray}
Re\frac{\epsilon_{xx}}{\epsilon_{0}}=1-\frac{1}{\eta}\frac{k_{x}}{k_{0}}\ .\end{eqnarray}
 Since $|\eta|\ll1$, there exists a broad region of $k_{x}$, \begin{eqnarray}
|\eta|k_{0}\ll|k_{x}|\ll k_{0},\end{eqnarray}
 such that the second term in (49) dominates and $q$ (see (41)) can
be approximated by \begin{eqnarray}
q=\left[k_{y}^{2}-\frac{k_{x}^{3}}{\eta k_{0}}\right]^{1/2}.\end{eqnarray}
 This expression, together with the aforementioned condition $q'\lesssim1/z$,
implies that the effective cutoff for $|k_{y}|$ is of order $1/z$,
whereas the cutoff for $|k_{x}|$ is of order $(|\eta|k_{0}/z^{2})^{1/3}$.
Substituting this value into (50), and returning to the physical quantities
$\omega$, $\tilde{\omega}_{p}$ and $v_{0}$, gives the condition
on $z$ which is necessary for the picture to be consistent: \begin{eqnarray}
|\frac{\omega-\tilde{\omega}_{p}}{v_{0}}|^{1/2}\left(\frac{v_{0}}{\tilde{\omega}_{p}}\right)^{3/2}\ll z\ll\frac{v_{0}}{|\omega-\tilde{\omega}_{p}|}.\end{eqnarray}
 Using the cutoffs for $k_{x}$ and $k_{y}$, one can now estimate
the integral in (44) with the following result, for $x'=x$: \begin{eqnarray}
\langle E^{2}(x,y,z)\rangle_{\omega}\simeq2\hbar\frac{\epsilon_{L}^{\prime\prime}}{|\epsilon_{0}(\omega)+1|^{2}}\Pi(\omega,T)\:\frac{|\omega-\tilde{\omega}_{p}|}{v_{0}z^{2}}.\end{eqnarray}
 Comparison between (53) and the corresponding result in equilibrium,
i.e., Eq.~(45) for $X=0$, reveals that under the conditions specified
above drift has a profound effect on the power spectrum of field fluctuations.
The ratio between (53) and the corresponding equilibrium quantity
is of order $\qquad\qquad$$|\omega-\tilde{\omega}_{p}|z/v_{0}$,
which due to (52) is much smaller than unity. Thus, our calculation
predicts a dip at the power spectrum for frequencies close to the
bulk plasmon frequency $\tilde{\omega}_{p}$.

The above estimate has been made under the condition given in Eq.~(46),
i.e., $\omega$ in Eq.~(53) cannot be too close to $\tilde{\omega}_{p}$.
In order to approach the immediate vicinity of the bulk plasmon frequency
one should replace the first inequality in (46) by the opposite one,
$|\omega-\tilde{\omega}_{p}|\ll\tilde{\omega}_{p}(\epsilon_{L}^{\prime\prime}/\epsilon'_{L})$.
Let us consider the extreme case $\omega=\tilde{\omega}_{p}$. For
this case Eq.~(36), with $\nu\rightarrow0$ and $|k_{x}|v_{0}\ll\tilde{\omega}_{p}$
gives \begin{eqnarray}
\epsilon_{xx}/\epsilon_{0}=1+2i\frac{\epsilon'_{L}}{\epsilon_{L}^{\prime\prime}}\frac{k_{x}v_{0}}{\tilde{\omega}_{p}}.\end{eqnarray}
 Note that this time it is essential to keep the imaginary part of
this ratio. Moreover, for the effect of drift to be significant, the
absolute value of the imaginary part must be much larger than unity,
i.e., \begin{eqnarray}
\frac{\epsilon_{L}^{\prime\prime}}{\epsilon'_{L}}\frac{\tilde{\omega}_{p}}{v_{0}}\ll|k_{x}|\ll\frac{\tilde{\omega}_{p}}{v_{0}}.\end{eqnarray}
 The quantity $q$, Eq.~(41), is now given by \begin{eqnarray}
q=\left[k_{y}^{2}+2ik_{x}^{2}\frac{\epsilon'_{L}}{\epsilon_{L}^{\prime\prime}}\frac{k_{x}v_{0}}{\tilde{\omega}_{p}}\right]^{1/2}.\end{eqnarray}
 and the effective cutoff for $k_{x}$ in the integral (44) is of
order $\left(\epsilon_{L}^{\prime\prime}\tilde{\omega}_{p}/\epsilon'_{L}v_{0}z^{2}\right)^{1/3}$.
Consistency with (55) requires $z\gg(v_{0}/\tilde{\omega}_{p})(\epsilon_{L}^{\prime\prime}/\epsilon'_{L})^{1/2}$.
The integral in (44) can now be estimated and, for $x'=x$, we obtain:
\begin{eqnarray}
\langle E_{x}^{2}(x,y,z)\rangle_{\omega}\simeq2\hbar\frac{(\epsilon_{L}^{\prime\prime})^{2}}{\epsilon'_{L}}\frac{1}{|\epsilon_{0}(\omega)+1|^{2}}\Pi(\omega,T)\frac{\tilde{\omega}_{p}}{v_{0}z^{2}}.\end{eqnarray}
 As expected, this value matches expression (53) at frequency $\omega$
such that $|\omega-\tilde{\omega}_{p}|\simeq\tilde{\omega}_{p}\epsilon_{L}^{\prime\prime}/\epsilon'_{L}$.
Eq.~(57) gives the minimum, at $\omega=\tilde{\omega}_{p}$, of the
dip in the spectral power.

Our consideration has been limited to the case when the main contribution
to the spectral power, i.e., to the integral in (44), comes from $k_{x}$
which satisfy the condition $|k_{x}|v_{0}\ll\omega$. This condition
imposes a restriction on $z$, namely the first inequality in (52)
(or its counterpart for frequencies very close to $\tilde{\omega}_{p}$).
For $z$ very close to the surface the integral in (44) is dominated
by large $k_{x}$, so that the condition $|k_{x}|v_{0}\ll\omega$
is violated. Considering values of $k_{x}$ larger than $\tilde{\omega}_{p}/v_{0}$
while still neglecting the thermal pressure requires the condition
$v_{0}>v_{T}$ which is better to be avoided. 

\section{Conclusion}

We have demonstrated the effect of carrier drift on thermal fluctuations
of the electric field. Only the rotationless part of the field has
been considered. This part dominates the short range correlations
in the bulk of the sample, as well as the fluctuations in the near
field sufficiently close to the sample surface. It has been shown
that drift can significantly affect the magnitude and the correlation
properties of the field fluctuations, especially at frequencies close
to the bulk plasma frequency. Our main goal was to discuss the effect
of drift in the simplest possible situation, making a number of idealizations
such as the limit of collisionless semiconductor plasma ($\nu\rightarrow0$)
or keeping (in Section 5) the ${\bf k}$-dependence due to drift only
in the component $\epsilon_{xx}$ of the permittivity tensor. The
latter simplification allowed us to avoid the intricate problem of
the additional boundary conditions which might be needed in the more
general case (see the book of Agranovich and Ginzburg in {[}17{]}).

This work is in some sense complementary to {[}16{]} where a collision
dominated transport regime was considered, i.e., it was assumed that
the scattering rate $\nu$ is much larger than the frequency $\omega$.
In this regime the linearized Ohmic current is simply ${\bf j}=\sigma_{0}{\bf E}+en{\bf v}_{0}$,
where $\sigma_{0}=e^{2}n_{0}/m\nu$ is the dc Drude conductivity.
Adding to this the diffusion current $-D\nabla n$, $D$ being the
diffusion coefficient, and eliminating $n$ with the help of the continuity
equation, one obtains instead of (13): \begin{eqnarray}
\epsilon(\omega,{\bf k})=\epsilon'_{L}\left[1+\frac{i/\tau_{M}}{\omega-{\bf k}\cdot{\bf v}_{0}+iDk^{2}}+i\frac{\epsilon_{L}^{\prime\prime}}{\epsilon'_{L}}\right],\end{eqnarray}
 where $\tau_{M}=\epsilon'_{L}/4\pi\sigma_{0}$ is the Maxwell relaxation
time. Neglecting $\epsilon_{L}^{\prime\prime}$, one recovers from
$\epsilon(\omega,{\bf k})=0$ the well known space charge waves, with
the dispersion relation $\omega={\bf k}\cdot{\bf v}_{0}-\frac{i}{\tau_{M}}-iDk^{2}$.
These waves can strongly influence current fluctuations in semiconductors
{[}12{]} and their impedance {[}22{]}. The surface counterpart of
such travelling waves can be excited at the semiconductor-vacuum boundary
{[}14{]} and can have a significant effect on the electromagnetic
field fluctuations near the surface {[}16{]}. It would be worthwhile
to further study the effect of these waves, under more general conditions
that those assumed in {[}16{]}.

One of the assumptions in the present work was that the spontaneous
random sources of the fluctuations were not affected by the drift
of the carriers. This is trivially so in the limit of collisionless
plasma when the lattice remains the only source of spontaneous fluctuations.
The assumption, though, can break down under more realistic conditions.
It should be possible to relax this assumption. Indeed, for the current
density fluctuations in the bulk, there is a well established theory
{[}12, 13{]} for the case when the carriers are way out of equilibrium
(\textquotedbl{}hot electrons\textquotedbl{}) and the corresponding
spontaneous random sources undergo a profound change. To our knowledge,
so far there is no extension of the theory to the case of surface
waves and their influence on the field fluctuations outside the sample.

It is clear that drift, via its effect on the electromagnetic field
fluctuations close to the sample surface, will influence also the
Casimir-Lifshitz forces {[}23{]}, as well as other related phenomena
(heat transfer, noncontact friction). Calculation of the forces involves
integration over frequencies, so that significant effect of drift
can be expected only if the main contribution to the integral comes
from the interval of frequencies in which the influence of drift on
field fluctuations is strong. The calculation of the Casimir-Lifshitz
forces in the presence of drift is beyond the scope of this paper.

Finally, let us emphasize that the effects considered in this work
are expected to occur only in materials with relatively low carrier
density (semiconductors, ionic conductors or other types of \textquotedbl{}bad
conductors\textquotedbl{}), where significant drift velocities can
be achieved without destroying the sample.

\section{Acknowledgement}

We acknowledge a useful discussion with I. Klich at an early stage
of this work. We are particularly grateful to L. Pitaevskii for carefully
reading the manuscript and making a number of valuable comments and
suggestions.

\pagebreak

\textbf{References}
\begin{enumerate}
\item E.M. Lifshitz and L.P. Pitaevskii, Statistical Physics, pt. II (Pergamon,
Oxford, 1980).
\item L.D. Landau and E.M. Lifshitz, Electrodynamics of Continuous Media
(Pergamon, Oxford, 1960).
\item M.L. Levin and S.M. Rytov, Theory of Equilibrium Thermal Fluctuations
in Electrodynamics (Nauka, Moscow 1967) (in Russian).
\item S.M. Rytov, Yu.A. Kravtsov and V.I. Tatarskii, Principles of Statistical
Radiophysics, Vol. 3, ch. 3 (Springer, Berlin, 1989).
\item K. Joulain, J.-P. Mulet, F. Marquier, R. Carminati and J.-J. Greffet,
Surface Science Reports \textbf{57}, 59 (2005).
\item A.I. Volokitin and B.N.J. Persson, Rev.\ Mod.\ Phys.\ \textbf{79},
1291 (2007).
\item M. Antezza, L.P. Pitaevskii, S. Stringari and V.B. Svetovoy, Phys.\ Rev.\ \textbf{A77},
022901 (2008).
\item G.L. Klimchitskaya, U. Mohideen and V.M. Mostepanenko, Rev.\ Mod.\
Phys. \textbf{81}, 1827 (2009).
\item M. Lax, Rev.\ Mod.\ Phys. \textbf{32}, 25, (1960).
\item M. Lax and P. Mengert, J.\ Phys.\
Chem.\ Solids \textbf{14}, 248 (1960).
\item L.E. Gurevich and B.I. Shapiro, Zh.\ Eksp.\ Teor.\ Fiz.\ \textbf{55},
1766 (1968) {[}Sov.\ Phys.\ JETP \textbf{28}, 931 (1969){]}.
\item Sh.M. Kogan and A.Ya. Shul'man, Zh. Eksp. Teor. Fiz. \textbf{56},
862 (1969) {[}Sov.\ Phys.\ JETP \textbf{29}, 467 (1969){]}.
\item S. V. Gantsevich, V.L. Gurevich and R. Katilyus, Zh.\ Eksp.\ Teor.\ Fiz.
\textbf{57}, 503 (1969) {[}Sov.\ Phys.\ JETP \textbf{30}, 276 (1970){]}.
\item G.S. Kino, IEEE Trans.\ Electron Devices, ED-17, 178 (1970).
\item B.G. Martin, J.J. Quinn and R.F. Wallis, Surf.\
Sci. \textbf{105}, 145 (1981).
\item A.M. Konin and B.I. Shapiro, Fiz.\ Tverd.\ Tela \textbf{14}, 2271
(1972) {[}Sov.\ Phys.\ - Sol.\ St.\ \textbf{14}, 1966 (1973){]}.
\item In general, Eq.(6) defines only an approximate branch of the spectrum,
with the retardation effects being neglected. For a thorough discussion
see V.M. Agranovich and V.L. Ginzburg, Crystal Optics with Spatial
Dispersion, and Excitons (Springer, 1984).
\item E.M. Lifshitz and L.P. Pitaevskii, Physical Kinetics (Pergamon, New
York, 1981).
\item We could have used the normally ordered function already in Sec. 4
but have preferred there the symmetrized version, as in Refs. {[}1-3{]}.
In any case, the difference appears only as an overall $T$-dependent
factor which has no bearing on the drift dependence of various quantities.
\item J. Schwinger et al., Classical Electrodynamics (Perseus Books, 1998).
\item C. Henkel, K. Joulain, R. Carminati and J.-J. Greffet, Opt. Commun.
\textbf{186}, 57 (2000).
\item L.E. Gurevich, I.V. Ioffe and B.I. Shapiro, Fiz. Tverd. Tela \textbf{11},
167 (1969). (Sov. Phys.-Sol. St. \textbf{11}, 124 (1969)).
\item The symmetrized version of the correlation functions must be used,
of course, while computing the Casimir-Lifshitz forces. In the absence
of drift, but in a non-equilibrium situation of a nonuniform temperature,
these forces has been studied in C. Henkel, K. Joulain, J.-P. Mulet
and J.-J. Greffet, J. Opt. \textbf{A4}, S109 (2002); M. Antezza, L.P.
Pitaevskii and S. Stringari, Phys. Rev. Lett. \textbf{95}, 113202
(2005); L.P. Pitaevskii, J. Phys. \textbf{A39}, 6665 (2006).
\end{enumerate}

\end{document}